\documentclass[aps,prd,reprint,a4paper,nofootinbib,superscriptaddress]{revtex4-2}

\usepackage{bbm}
\usepackage{amsmath}
\usepackage{graphicx}
\usepackage{float}
\usepackage{amssymb}
\usepackage{subfigure}
\usepackage{amssymb}
\usepackage{hyperref}
\usepackage{epstopdf}

\usepackage[utf8]{inputenc}
\hypersetup{
    colorlinks=true,
    linkcolor=red,
    citecolor=blue,
}
\usepackage{color}
\usepackage[T1]{fontenc}
\usepackage{txfonts}
\usepackage{mathrsfs}
\usepackage{latexsym}
\usepackage{epsfig}
\usepackage{epstopdf}
\usepackage{epstopdf}
\usepackage{graphicx}
\usepackage{amssymb}
\usepackage{amsmath}
\usepackage{dcolumn}
\usepackage{bm}
\usepackage{color}
\usepackage{comment}
\usepackage{xcolor}
\usepackage{dsfont}

\newcommand{\lcr}{l_{\rm cr}}
\newcommand{\Mdot}{\dot M}
\newcommand{\rin}{r_{\rm in}}
\newcommand{\Msun}{M_\odot}

\begin{document}
\raggedbottom

\title{Hidden by Exponential Screening: The Invisibility of the Dymnikova Black Hole's de Sitter Core to Accretion Observables}

\author{Orhan~Donmez}
\email{orhan.donmez@aum.edu.kw}
\affiliation{College of Engineering and Technology, American University of the Middle East, Egaila 54200, Kuwait}

\author{Sushant~G.~Ghosh}\email{sghosh2@jmi.ac.in}
\affiliation{Centre for Theoretical Physics,
Jamia Millia Islamia, New Delhi 110025, India}
\affiliation{Center for Theoretical Physics, Khazar University, 41 Mehseti Street, Baku, AZ1096, Azerbaijan}
\affiliation{Astrophysics and Cosmology Research Unit,
School of Mathematics, Statistics and Computer Science, University of KwaZulu-Natal, Private Bag 54001, Durban 4000, South Africa}

\author{Farruh~Atamurotov}
\email{atamurotov@yahoo.com}
\affiliation{New Uzbekistan University, Movarounnahr street 1, Tashkent 100000, Uzbekistan}

\author{G. Mustafa}
\email{gmustafa3828@gmail.com }
\affiliation{Department of Physics, Zhejiang Normal University,
Jinhua 321004, China}
\affiliation{Research Center of Astrophysics and Cosmology, Khazar University, Baku, AZ1096, 41 Mehseti Street, Azerbaijan}

\date{\today}

\begin{abstract}
The Dymnikova spacetime replaces the Schwarzschild singularity with a de Sitter core, and its lapse deviates from Schwarzschild by a factor decaying as \(e^{-r^3/(2l_{\rm cr}^2M)}\). Whether this screening renders the core invisible to accretion has remained an open quantitative question. We address it with general-relativistic hydrodynamic simulations of Bondi--Hoyle--Lyttleton accretion onto Dymnikova black holes for six values of the critical length, \(0.1M\le l_{\rm cr}\le 1.1M\), the last being \(97\%\) of the extremal value. All runs share identical gas parameters and resolution, so any difference in the flow must originate from the geometry. The dynamically relevant quantity is the gravitational acceleration, whose deviation reaches \(12\%\) at the inner boundary for the most strongly corrected model, yet the flow responds only locally: the post-shock density drops by \(10\%\) in the innermost zone and the quiet-phase accretion rate by \(0.12\%\), with no effect beyond \(r\simeq2.7M\). Because runs differing by as little as \(\sim 10^{-10}\) in the metric decorrelate completely once the shock-cone instability saturates, we compare models statistically using ensembles of independent realizations. The time-averaged accretion rate is then independent of \(l_{\rm cr}\) to within \(0.8\%\) (\(2\sigma\)) at all radii; the variability amplitude, shock-cone opening angle (\(1.077\pm0.003\) rad), and stagnation radius (\(26.8\pm0.3M\)) vary only within the realization scatter; the propagation lags (\(75.5\pm1.0M\) and \(31.6\pm0.4M\)) agree to within \(1.6\%\); and no power spectrum carries a detectable core signature. Accretion onto a Dymnikova black hole is therefore observationally indistinguishable from accretion onto a Schwarzschild black hole: timing observations test the shock-cone mechanism, not the regular core. More broadly, ensemble-based statistical comparisons are essential before differences between simulated accretion flows in different spacetimes can be interpreted as physical.

\end{abstract}


\maketitle

\date{\today}

\section{Introduction}
The idea that classical black-hole singularities might be resolved by
quantum-gravity effects has motivated a broad family of regular
black-hole geometries, in which the central curvature singularity is
replaced by a regular core. Among the earliest and most influential of
these constructions is the Dymnikova spacetime, proposed more than three
decades ago as a vacuum nonsingular black hole in which the
Schwarzschild singularity is replaced by a de Sitter core while
asymptotic flatness is preserved at large distances
\cite{Dymnikova:1992ux}. The geometry is generated by an effective
stress--energy tensor that violates the strong energy condition in the
core but remains physically acceptable as an effective description, and
it has since become a standard reference point for singularity-free
black holes. The Dymnikova solution has been generalized to higher
dimensions in both general relativity and Rastall theory, where the
photon-sphere radii are found to depend on the number of extra
dimensions \cite{Paul:2023pqn}. It has also been shown to arise from an
infinite tower of higher-curvature corrections, establishing a
connection between regular black holes and quasi-topological gravity
\cite{Konoplya:2024kih}. The question of how such regular black holes
might form dynamically has been addressed through models of
gravitational collapse of baryonic matter into a de Sitter core, with
the transition to exotic matter accompanied by a detectable
electromagnetic radiation signature \cite{Vertogradov:2025aok}. The
Dymnikova geometry has also been extended to include surrounding matter
distributions: generalized solutions immersed in perfect fluid dark
matter and a cloud of strings have been analyzed, with the Hawking
temperature, particle dynamics, and quasi-periodic oscillation
frequencies computed as functions of the additional parameters \cite{Ahmed:2026vce}. The Dymnikova solution has been generalized to rotating cases, where the shadow and ergosphere are found to be sensitive to the charge parameter \cite{Ghosh:2020ece}.

A second, independent line of reasoning leads to the same functional
form. In asymptotically safe gravity, the Newton coupling is promoted
to a running quantity determined by a renormalization-group cutoff.
When the cutoff is fixed iteratively as a functional of the effective
energy density, the resulting lapse function converges precisely to the
Dymnikova metric, with the de Sitter core radius related to the critical
length scale of the quantum correction \cite{Platania:2019kyx,
Bonanno:2000ep}. The Bonanno--Reuter construction is among the earliest
regular black-hole metrics derived in this framework, in which the
running Newton coupling eliminates the ultraviolet singularity and
produces a nonsingular core \cite{Bonanno:2000ep}. A comprehensive review
of black holes in asymptotically safe gravity, including the derivation
of the Dymnikova metric from functional renormalization group
improvements, is given in \cite{Platania:2023srt}. The Dymnikova spacetime
is therefore doubly motivated: it is both a phenomenological regular
black hole and a concrete manifestation of
quantum-gravity-inspired corrections to the Schwarzschild geometry. 

The defining feature of this geometry is the exponential screening of
the core. Writing the lapse as \(f=f_{\rm S}+\delta f\), the relative
deviation from Schwarzschild behaves as
\(\delta f/f_{\rm S}\propto e^{-r^3/(2l_{\rm cr}^2M)}\), so that the
correction is confined to a thin shell around the would-be horizon and
becomes astronomically small at the photon sphere and beyond. This
screening has direct consequences for black-hole spectroscopy.
Gravitational quasinormal modes of the Dymnikova black hole show
systematic but modest deviations from the Schwarzschild spectrum: the
real oscillation frequency decreases as \(l_{\rm cr}\) increases, and
the damping rate becomes smaller, implying longer-lived modes, with the
spectrum smoothly approaching Schwarzschild in the large-\(l_{\rm cr}\)
limit \cite{Lutfuoglu2026AnnPhys}. Earlier studies reached similar
conclusions for the fundamental modes, with only higher overtones
responding strongly to the near-horizon deformation
\cite{Konoplya:2023aph}. Grey-body factors and absorption cross-sections of gravitational perturbations remain close to their Schwarzschild values, with deviations concentrated at low frequencies and controlled by the core size \cite{Dubinsky:2025nxv}. The Hawking radiation has also been analyzed in
detail: grey-body factors for Standard Model fields and gravitons were
computed, and the main effect was found to be not a dramatic change in
transmission probabilities but a rapid decrease of the Hawking
temperature, which strongly suppresses the total luminosity near the
extremal remnant \cite{Skvortsova:2026mwz}. Event Horizon Telescope
constraints on Dymnikova-type regular black holes in higher dimensions
show that the shadow size remains compatible with observations for a
broad range of parameters \cite{Errehymy:2026ftm}. The same screening
suggests that electromagnetic observables tied to the strong-field
region may be similarly insensitive to the core.

Accretion offers an alternative and complementary probe.
Bondi--Hoyle--Lyttleton (BHL) accretion, in which a black hole moves
through a gaseous medium or is immersed in a wind, is the natural regime
for isolated black holes and for low-angular-momentum inflows in
galactic nuclei \cite{Edgar:2004mk}. General-relativistic hydrodynamic
simulations of BHL flows around Schwarzschild black holes have
established that the shock cone that forms downstream of the accretor
becomes unstable and produces broadband accretion-rate variability, with
quasi-periodic oscillations naturally generated in the cone
\cite{Donmez:2010sx}.

More recently, BHL accretion has been developed into a systematic probe
of strong-field geometry in modified spacetimes. Simulations around
Kerr--Sen black holes show that the charge-related deformation shifts
the characteristic shock-cone oscillation frequencies and produces
near-resonant harmonic structures \cite{Donmez:2026ntl}. In the
Johannsen--Psaltis metric, the shock-cone opening angle and the
resulting quasi-periodic oscillations respond sensitively to the
deviation parameter \cite{Donmez:2026yvm}. In Weyl conformal gravity, the
curvature term acts at the accretion radius and can drive flip-flop
oscillations or destroy the cone entirely, producing QPOs that would be
detectable within seconds for a bright Galactic source
\cite{Donmez:2026tdt}. For quantum-corrected black holes, BHL
simulations show that quantum corrections suppress peak QPO amplitudes
near the horizon, shift the QPO-generating regions outward, and preserve
recognizable resonance patterns such as the 3:2 frequency ratio
\cite{Donmez2025EPJC,Mustafa:2025mkc}. These studies collectively demonstrate that the
sensitivity of BHL accretion to a modified geometry depends on how far
the deviation extends beyond the horizon, and on whether it acts at the
accretion radius or only in the deep strong-field region.

Accretion onto regular black holes specifically has also begun to be
explored. The accretion mechanism for regular black holes with
asymptotically Minkowski cores and for improved Schwarzschild black
holes has been analyzed within a general isotropic-fluid framework,
identifying the critical parameters that control the radial velocity,
energy density, sound speed, and mass accretion rate
\cite{Murtaza:2024nfy}. Comparative studies of Bondi and Novikov–Thorne accretion around regular black holes—including the Dymnikova solution—find that the regularization parameter shifts the sonic radius outward and produces systematically higher luminosities than Schwarzschild \cite{Capozziello:2025ycu,2025hasb.confE..36G}. These studies, however, are restricted to
stationary or spherically symmetric configurations. The present work
extends this line of inquiry to time-dependent, non-axisymmetric BHL
flows, where the nonlinear dynamics of the shock cone and the
stochasticity of the resulting variability introduce additional layers
of complexity.

There is nevertheless good reason to expect a signature. The gas
dynamics are governed by metric derivatives through the Christoffel
symbols, and the relative deviation of the gravitational acceleration
\(\delta g/g_{\rm S}\) exceeds \(\delta f/f_{\rm S}\) by a factor of 3--10 or more near the black hole (Sec.~\ref{subsec:screening}). Stationary studies show that a de Sitter core can alter the accretion flow, either reducing the infall
velocity or enhancing it \cite{Neves:2019ywx}. For the most strongly
corrected Dymnikova configurations, the gravitational pull at the inner
boundary is reduced by more than ten per cent. Whether this local
weakening leaves any imprint on global observables---mean rate,
variability amplitude, shock-cone geometry, propagation lags, or power
spectrum---requires controlled numerical experiments that separate
physical effects from the intrinsic stochasticity of the flow. Existing
regular black hole accretion studies are restricted to stationary Bondi
or Novikov--Thorne configurations \cite{Capozziello:2025ycu}, whereas
relativistic BHL simulations have established that the shock cone
produces stochastic variability and flip-flop oscillations inaccessible
to analytic treatment \cite{Donmez:2010sx}.

This paper reports such an investigation. We perform general-relativistic
hydrodynamic simulations of BHL accretion onto Dymnikova black holes for
six values of $l_{\rm cr}$, spanning from geometries that are
Schwarzschild to machine precision to the near-extremal configuration.
All runs use identical gas parameters and resolution, so any difference
in the flow must originate from the geometry. Because runs differing by
as little as $\sim 10^{-10}$ in the metric decorrelate completely once the
instability saturates, we compare models statistically, using ensembles
of independent realizations to establish the intrinsic scatter. This
allows us to place quantitative bounds on the sensitivity of accretion
to the de Sitter core.

The paper is organized as follows. Section~\ref{sec:dymnikova} reviews the Dymnikova
geometry, its renormalization-group origin, and the exponential
screening of the correction, and derives the bounds on the metric
deviation at the radii relevant to accretion. Section~\ref{sec:setup} describes the
numerical setup, the initial and boundary conditions, and the diagnostic
and statistical methods. Section~\ref{sec:results} presents the results: the shock-cone
morphology and the local response to the core; the evolution of the
accretion rate and the sensitivity of the flow to perturbations; the
time-averaged rate and variability amplitude; the inward propagation of
fluctuations; and the power spectra, including a critical assessment of
the significance of quasi-periodic features. Section~\ref{sec:astro} discusses the
astrophysical implications, including the time scales for observed
black-hole sources and the conditions under which a core could in
principle be revealed. Section~\ref{sec:conclusions} summarizes our conclusions.

\section{The Dymnikova regular black hole}
\label{sec:dymnikova}
A central challenge in classical general relativity is that gravitational
collapse terminates in a curvature singularity, where the theory loses
predictive power. This breakdown is widely regarded as a signal that
Einstein's equations must be modified by quantum-gravitational effects
in the strong-field regime, with the expectation that such corrections
replace the singularity by a regular core of bounded curvature. Among
the earliest and most influential realizations of this idea is the
Dymnikova spacetime, proposed more than three decades ago as a vacuum
nonsingular black hole in which the Schwarzschild singularity is
replaced by a de Sitter core while asymptotic flatness is preserved at
large distances \cite{Dymnikova:1992ux}. 
The Dymnikova spacetime is static and spherically symmetric,
\begin{align}
ds^2&=-f(r)\,dt^2+\frac{dr^2}{f(r)}+r^2\left(d\theta^2+\sin^2\theta\,d\phi^2\right),\nonumber\\
f(r)&=1-\frac{2m(r)}{r},
\label{eq:metric}
\end{align}
with the mass function~\cite{Dymnikova:1992ux}
\begin{equation}
m(r)=M\left(1-e^{-r^3/r_0^3}\right),
\label{eq:massfn}
\end{equation}
where \(M\) is the Arnowitt--Deser--Misner mass and \(r_0\) the radius of
the de Sitter core. For \(r\ll r_0\), \(f\simeq1-2Mr^2/r_0^3\), which
is the de Sitter form, and for \(r\gg r_0\) the Schwarzschild metric is
recovered. The geometry is regular everywhere, with all curvature
invariants finite at \(r=0\), at the price of an effective
stress--energy tensor that violates the strong energy condition in the
core~\cite{Dymnikova:1992ux}. The Dymnikova solution has been generalized
to higher dimensions in both general relativity and Rastall theory,
where the photon-sphere radii depend on the number of extra
dimensions~\cite{Paul:2023pqn}, and its thermodynamics and quasinormal spectra have been extended to higher dimensions~\cite{Macedo2024} and to a black hole surrounded by a cloud of strings~\cite{SantosBarbosa2026}.

The same form follows from asymptotic safety~\cite{Platania:2019kyx,Bonanno:2000ep}. Replacing Newton's constant in the Schwarzschild lapse by a running coupling,
\begin{equation}
f(r)=1-\frac{2M}{r}\frac{G(r)}{G_0},\qquad G(r)=\frac{G_0}{1+G_0k^2(r)/g_*},
\label{eq:running}
\end{equation}
where $g_*$ is the non-Gaussian fixed point and $k(r)$ an infrared cutoff that vanishes at infinity, produces an effective anisotropic fluid with energy density and pressure
\begin{equation}
\rho_{\rm eff}=\frac{MG'(r)}{4\pi r^2G(r)},\qquad p_{\rm eff}=-\frac{MG''(r)}{8\pi rG(r)},
\label{eq:effective}
\end{equation}
which may be interpreted as vacuum polarization. When the cutoff is updated iteratively as a functional of $\rho_{\rm eff}$, the sequence of metrics converges to~\cite{Platania:2019kyx,mainsol}
\begin{equation}
f(r)=1-\frac{2M}{r}\left(1-e^{-r^3/(2\lcr^2M)}\right),
\label{eq:dymnikova}
\end{equation}
i.e., to the Dymnikova metric with $r_0^3=2\lcr^2M$. The critical length $\lcr$ characterizes the scale of the quantum correction, and $\lcr=0$ gives back Schwarzschild. Two horizons $r_-<r_+$ exist for $\lcr<\lcr^{\rm ext}$; they merge at the extremal value, determined by $f=f'=0$, which we find to be
\begin{equation}
\lcr^{\rm ext}=1.1379M,\qquad r_{\rm ext}=1.702M,
\label{eq:extremal}
\end{equation}
and for larger $\lcr$ the spacetime describes a horizonless, nonsingular compact object. In the following we set $M=1$.

\subsection{Exponential screening of the correction}
\label{subsec:screening}

Writing $f=f_{\rm S}+\delta f$, with $f_{\rm S}=1-2M/r$, and defining $x\equiv r^3/(2\lcr^2M)$, the relative deviation of the lapse takes the exact closed form
\begin{equation}
\frac{\delta f}{f_{\rm S}}=\frac{2M\,e^{-x}}{r-2M}.
\label{eq:df}
\end{equation}
The dynamics of the gas, however, are governed by derivatives of the metric, which also control the tidal forces acting on extended bodies falling into a Dymnikova black hole~\cite{Macedo2026}. A convenient measure is the proper acceleration of static observers, $g=f'/(2\sqrt f)$, which sets the effective gravitational pull in the momentum equation. Using $f'_{\rm S}=2M/r^2$ and $f'=f'_{\rm S}-(2M/r^2)(1+3x)e^{-x}$, its relative deviation is, to first order,
\begin{equation}
\frac{\delta g}{g_{\rm S}}=-e^{-x}\left[1+3x+\frac{M}{r-2M}\right].
\label{eq:dg}
\end{equation}
The factor $1+3x$, which arises from $m'(r)$, makes $|\delta g/g_{\rm S}|$ larger than $|\delta f/f_{\rm S}|$ by a factor of $3$--$10$ or more close to the black hole. The correction to the gravitational pull is negative: the de Sitter core weakens gravity.

For $x>2/3$, which holds everywhere outside $r=2.3M$ for all black-hole configurations, Eqs.~(\ref{eq:df}) and (\ref{eq:dg}) decrease monotonically with $r$ at fixed $\lcr$ and increase monotonically with $\lcr$ at fixed $r$. Their values for the extremal configuration therefore bound the deviation for every Dymnikova black hole. Evaluating them for $\lcr=\lcr^{\rm ext}$ gives
\begin{align}
r=3M:&\ \left|\delta f/f_{\rm S}\right|<5.9\times10^{-5},\ \left|\delta g/g_{\rm S}\right|<1.0\times10^{-3},\nonumber\\
r=6M:&\ \left|\delta f/f_{\rm S}\right|<3.0\times10^{-37},\ \left|\delta g/g_{\rm S}\right|<1.5\times10^{-34}.
\label{eq:bounds}
\end{align}
Hence, the photon sphere, the shadow, and above all the ISCO and the region outside it are, for all practical purposes, those of a Schwarzschild black hole. For $\lcr=1.1M$, Eq.~(\ref{eq:dg}) gives $\delta g/g_{\rm S}=-12.7\%$ at $\rin$ to first order (the exact value is $-12.4\%$). For the same model, the photon sphere moves inward by only $5.2\times10^{-4}M$ and the shadow radius changes by $1.4\times10^{-5}$ in relative terms, far below the precision of horizon-scale imaging~\cite{EHT2019M87VI, Psaltis2020,EHT2022SgrAVI}, consistent with the weak shadow constraints obtained for Dymnikova-type black holes~\cite{Errehymy:2026ftm,Macedo2025}. The deviation can only be appreciable inside $r\simeq3M$.

\begin{figure*}[t]
\centering
\includegraphics[width=\textwidth]{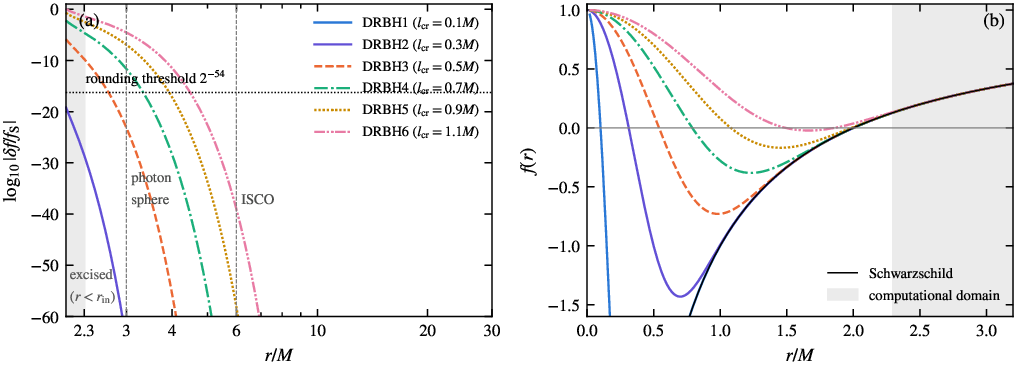}
\caption{(a) Relative deviation of the lapse function from Schwarzschild, Eq.~(\ref{eq:df}), for the six models of Table~\ref{tab:models}. The grey band marks the excised region $r<\rin=2.3M$, and the horizontal dotted line the double-precision rounding threshold $2^{-54}$; models DRBH1 and DRBH2 lie below it everywhere in the domain. The dashed vertical lines mark the Schwarzschild photon sphere and ISCO. (b) Lapse function $f(r)$ of the same models compared with Schwarzschild (black); the grey band marks the inner part of the computational domain.}
\label{fig:metric}
\end{figure*}

Figure~\ref{fig:metric}(a) shows $\delta f/f_{\rm S}$ for the models used in this work, and Fig.~\ref{fig:metric}(b) the lapse function in the interior. A practical consequence of Eq.~(\ref{eq:df}) is that in double-precision arithmetic the factor $1-e^{-x}$ rounds exactly to unity when $e^{-x}<2^{-54}$, i.e., for $r>r_\epsilon\equiv(108\ln2\,\lcr^2M)^{1/3}$. For $\lcr=0.1M$ and $0.3M$ we find $r_\epsilon=0.91M$ and $1.89M$, both inside the inner boundary of our grid. These two configurations are therefore bitwise Schwarzschild inside the computational domain. We use this property in Sec.~\ref{subsec:chaos} to calibrate the intrinsic variability of the flow.

\begin{table}[t]
\caption{Dymnikova models. $r_\pm$ are the horizon radii; $\delta f/f_{\rm S}$ and $\delta g/g_{\rm S}$ [Eqs.~(\ref{eq:df}) and (\ref{eq:dg})] are evaluated at the inner boundary $\rin=2.3M$; $r_\epsilon$ is the radius beyond which the correction is lost to rounding in double precision. $\delta g/g_{\rm S}$ is the first-order value of Eq.~(\ref{eq:dg}).}
\label{tab:models}
\setlength{\tabcolsep}{2.5pt}
\begin{ruledtabular}
\begin{tabular}{lcccccc}
Model & $\lcr/M$ & $r_-/M$ & $r_+/M$ & $\delta f/f_{\rm S}$ & $|\delta g/g_{\rm S}|$ & $r_\epsilon/M$\\
\hline
DRBH1 & 0.1 & 0.1013 & 2.0000 & $4\times10^{-264}$ & $1\times10^{-261}$ & 0.91\\
DRBH2 & 0.3 & 0.3128 & 2.0000 & $3\times10^{-29}$ & $9\times10^{-28}$ & 1.89\\
DRBH3 & 0.5 & 0.5398 & 2.0000 & $1.8\times10^{-10}$ & $2.1\times10^{-9}$ & 2.66\\
DRBH4 & 0.7 & 0.7895 & 1.9994 & $2.7\times10^{-5}$ & $1.7\times10^{-4}$ & 3.32\\
DRBH5 & 0.9 & 1.0790 & 1.9839 & $3.7\times10^{-3}$ & $1.5\times10^{-2}$ & 3.93\\
DRBH6 & 1.1 & 1.4895 & 1.8599 & $4.4\times10^{-2}$ & $1.3\times10^{-1}$ & 4.49\\
\end{tabular}
\end{ruledtabular}
\end{table}

\section{Numerical setup and analysis methods}
\label{sec:setup}

\subsection{General relativistic hydrodynamics}
\label{subsec:grhd}

We solve the equations of general relativistic hydrodynamics on the fixed Dymnikova background in the equatorial plane, neglecting the self-gravity of the gas, magnetic fields, and viscosity. In the $3+1$ formalism, the equations are written in conservative form~\cite{Banyuls1997,Font2000,Rezzolla2013},
\begin{equation}
\partial_t\mathbf{U}+\partial_r\mathbf{F}^r+\partial_\phi\mathbf{F}^\phi=\mathbf{S},
\label{eq:grhd}
\end{equation}
with conserved variables, fluxes, and sources
\begin{equation}
\mathbf{U}=
\begin{pmatrix} D \\ S_j \\ \tau \end{pmatrix}
=\sqrt{\gamma}
\begin{pmatrix}
W\rho \\
\rho h W^2 v_j \\
\rho h W^2 - P - W\rho
\end{pmatrix},
\label{eq:cons}
\end{equation}
\begin{equation}
\mathbf{F}^i=\alpha
\begin{pmatrix}
\tilde v^i D \\
\tilde v^i S_j+\sqrt{\gamma}\,P\,\delta^i_j \\
\tilde v^i \tau+\sqrt{\gamma}\,P\,v^i
\end{pmatrix},\qquad \tilde v^i=v^i-\frac{\beta^i}{\alpha},
\label{eq:flux}
\end{equation}
\begin{equation}
\mathbf{S}=\alpha\sqrt{\gamma}
\begin{pmatrix}
0 \\
T^{\mu\nu} g_{\nu\sigma}\Gamma^{\sigma}_{\mu j} \\
T^{\mu 0}\partial_\mu \alpha-\alpha\, T^{\mu\nu}\Gamma^{0}_{\mu\nu}
\end{pmatrix}.
\label{eq:source}
\end{equation}
Here $\rho$, $P$, $\epsilon$, and $h=1+\epsilon+P/\rho$ are the rest-mass density, pressure, specific internal energy, and specific enthalpy; $v^i$ is the three-velocity measured by the Eulerian observer; $W=(1-\gamma_{ij}v^iv^j)^{-1/2}$ is the Lorentz factor; $T^{\mu\nu}$ is the perfect-fluid stress--energy tensor; and $\Gamma^\sigma_{\mu\nu}$ are the Christoffel symbols~\cite{MTW1973}. For the metric~(\ref{eq:metric}),
\begin{equation}
\alpha=\sqrt{f},\qquad \beta^i=0,\qquad
\gamma_{ij}={\rm diag}\left(f^{-1},\,r^2,\,r^2\sin^2\theta\right).
\label{eq:3p1}
\end{equation}
The metric enters the evolution through $\alpha$, $\sqrt\gamma$, and, most importantly, through the Christoffel symbols in the source terms, which contain $f'$. This is why the response of the flow is controlled by $\delta g/g_{\rm S}$ rather than by $\delta f/f_{\rm S}$ (Sec.~\ref{subsec:screening}). The system is closed with the ideal-gas equation of state $P=(\Gamma-1)\rho\epsilon$. Equation~(\ref{eq:grhd}) is solved with a high-resolution shock-capturing scheme based on an approximate Riemann solver, with the time step set by the Courant condition~\cite{Donmez2004,Donmez:2010sx}.

\subsection{Initial and boundary conditions}
\label{subsec:initial}

The domain is initially filled with a uniform gas of density $\rho_\infty=1$, sound speed $C_\infty=0.1$, and adiabatic index $\Gamma=4/3$, moving with asymptotic speed $V_\infty=0.2$ along the $x$ direction,
\begin{equation}
v^r=\sqrt{\gamma^{rr}}\,V_\infty\cos\phi,\qquad
v^\phi=-\sqrt{\gamma^{\phi\phi}}\,V_\infty\sin\phi,
\label{eq:initvel}
\end{equation}
so that $\phi=\pi$ is the upstream and $\phi=0$ the downstream direction. The accretion radius is $r_{\rm acc}=2M/(V_\infty^2+C_\infty^2)\simeq40M$~\cite{Edgar:2004mk}. The gas parameters are the same in all runs, so any difference in the flow must originate from the geometry. The six models are listed in Table~\ref{tab:models}; all of them are black holes, with $\lcr<\lcr^{\rm ext}$.

The equations are evolved in $r\in[2.3M,100M]$, $\phi\in[0,2\pi]$, on a uniform grid with $N_r\times N_\phi=1024\times256$ zones ($\Delta r=0.095M$, $\Delta\phi=0.0245$~rad). The inner boundary $\rin=2.3M$ lies outside the outer horizon of every model. At $\rin$, an outflow condition lets the gas fall into the black hole. At the outer boundary, gas is injected with the initial state in the upstream half ($\pi/2<\phi<3\pi/2$), while outflow conditions are used downstream. The $\phi$ direction is periodic~\cite{Donmez:2010sx,Donmez2024JCAP,Donmez2025EPJC}.

\subsection{Diagnostics}
\label{subsec:diagnostics}

The mass accretion rate is computed from the radial flux of the conserved density through circles of fixed radius, as in our earlier work~\cite{Donmez:2010sx,Donmez2024JCAP},
\begin{equation}
\Mdot(r,t)=-\int_0^{2\pi}\alpha\,D\,v^r\,d\phi=-\int_0^{2\pi}\alpha\sqrt{\gamma}\,W\rho\,v^r\,d\phi,
\label{eq:mdot}
\end{equation}
evaluated at $r=2.3M$ (inner boundary), $6.11M$ (close to the Schwarzschild ISCO), and $12M$. It is recorded every $\simeq0.036M$ from $t=3000M$, by which time the shock cone has formed, until the end of each run ($t_{\rm end}=1.93$--$2.39\times10^4M$). For all statistical comparisons we use the common window $t\in[5000,19300]M$, of length $T=1.43\times10^4M$ ($0.70$~s for $M=10\Msun$), which starts after the instability has saturated (Sec.~\ref{subsec:chaos}) and ends when the shortest run finishes. This gives every model the same spectral resolution, $\Delta f=1/T=1.42$~Hz, and the same noise properties. The time series are averaged into bins of $1M$ before the analysis. The time-averaged $\Mdot$ differs by $\simeq3\%$ between the three radii; since we always compare models at fixed radius, this offset, which is the same in all runs, does not affect our conclusions. We also use one-dimensional cuts of the density and velocity at the final time of each run: along the downstream axis $\phi=0$ (written out at every second radial zone, $\Delta r=0.19M$), and along $\phi$ at $r=2.66M$ and $6.11M$.

\subsection{Statistical methods}
\label{subsec:stats}

\subsubsection{Time-domain statistics} 
For each run and radius we compute the mean, the fractional rms $\sigma/\langle\Mdot\rangle$, and the skewness and excess kurtosis. Because the series are strongly autocorrelated, we estimate the integrated autocorrelation time $\tau_{\rm int}$ and the uncertainty of the mean as $\sigma\sqrt{2\tau_{\rm int}/T}$; we verified it against a moving-block bootstrap with blocks of $1000M$ and adopt the larger of the two. The relative uncertainty of $\sigma$ is estimated as $\sqrt{\tau_{\rm int}/T}$.

\subsubsection{Separation between runs} 
To quantify how two runs $a$ and $b$ decorrelate, we use
\begin{equation}
\mathcal{D}(t)=\frac{\left\langle(\Mdot_a-\Mdot_b)^2\right\rangle_{100M}^{1/2}}{\left(\sigma_a^2+\sigma_b^2\right)^{1/2}},
\label{eq:sep}
\end{equation}
where the average is over $100M$ windows and $\sigma_a$ and $\sigma_b$ are the rms values in the saturated phase. For identical runs $\mathcal{D}=0$, and for statistically independent ones $\mathcal{D}\to1$.

\subsubsection{Power spectra} 
We compute rms-normalized periodograms $I_j=(2\Delta t/N)|X_j|^2/\langle\Mdot\rangle^2$ of the $N=14300$ binned points~\cite{vanderKlis1989}. Each periodogram is fitted with a bending power law plus constant,
\begin{equation}
S(f)=\frac{A f^{-\alpha_1}}{1+(f/f_b)^{\alpha_2-\alpha_1}}+C_0,
\label{eq:bpl}
\end{equation}
by minimizing the Whittle statistic $C=2\sum_j[\ln S_j+I_j/S_j]$, appropriate for $\chi^2_2$-distributed periodogram ordinates~\cite{Vaughan2010}. The fit uses all $211$ frequencies between $1.42$ and $300$~Hz. Candidate QPOs are searched for in two ways. First, the ratio $2I_j/S_j$ is compared with the $\chi^2_2$ distribution, and with $\chi^2_{2m}$ after averaging $m=2$--$4$ adjacent bins, including the number of trial frequencies in the false-alarm probability~\cite{Vaughan2005}. Second, a Lorentzian
\begin{equation}
L(f)=\frac{R}{\pi/2+\arctan(2Q)}\,\frac{\Delta}{(f-f_0)^2+\Delta^2},\qquad\Delta=\frac{f_0}{2Q},
\label{eq:lor}
\end{equation}
normalized so that $R$ is its fractional rms$^2$ over $f>0$, is added to the continuum. Following the usual definition of a QPO~\cite{Belloni2002}, we restrict this component to $Q\ge2$ and $2\le f_0\le200$~Hz. The improvement $\Delta C$ is calibrated with $600$ simulated periodograms drawn from the best-fitting continuum and analyzed with exactly the same search~\cite{TimmerKonig1995,Protassov2002}. Empirical $p$ values are quoted; the $3\sigma$ ($99.73\%$) threshold, which is poorly sampled by 600 draws (only $\simeq 1.6$ are expected to exceed it), is obtained from a scaled $\chi^2$ distribution matched to the first two moments of the simulated $\Delta C$, which reproduces the empirical $99\%$ quantile. Since the five runs DRBH1, DRBH3--DRBH6 are statistically independent (Sec.~\ref{subsec:chaos}), we also average their periodograms; the ensemble mean of $n=5$ periodograms is fitted with $C=2n\sum_j[\ln S_j+\bar I_j/S_j]$. Finally, the variance in five frequency bands is compared between runs with an $F$ test against the mean of the other four runs.

\subsubsection{Cross-spectra} 
Time lags between radii are computed from the phase of the cross spectrum, averaged over $12$ Hann-windowed segments of $2048M$ with $50\%$ overlap, and their uncertainties follow from the coherence~\cite{BendatPiersol2010,Nowak1999}. A positive lag $\tau_{a\to b}$ means that the variability at radius $b$ follows that at radius $a$.

\section{Results}
\label{sec:results}

\begin{figure*}[t]
\centering
\includegraphics[width=\textwidth]{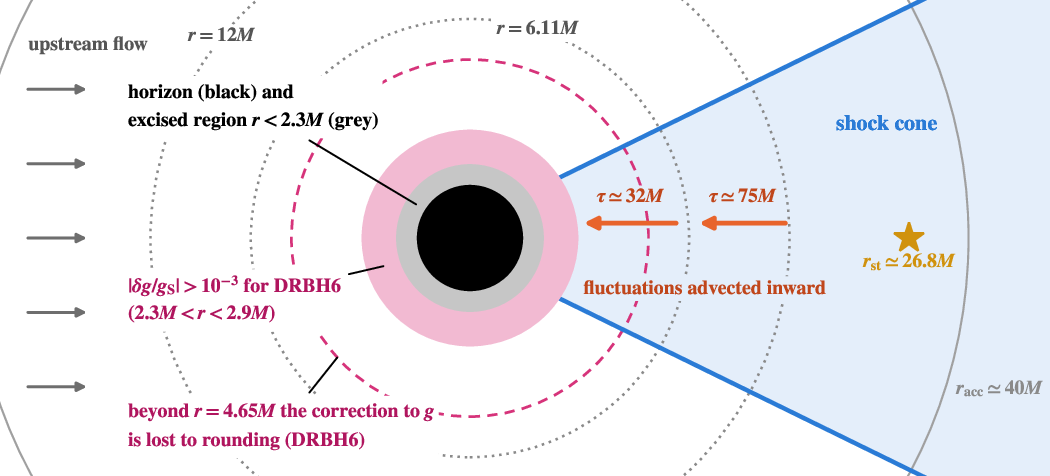}
\caption{Schematic view of the BHL flow in the equatorial plane, with the radius compressed as $\ln(r/1.4M)$ (not to scale). Gas enters from the left and is focused into the downstream shock cone (blue), whose stagnation point is at $r_{\rm st}\simeq26.8M$ (star). Accretion-rate fluctuations are advected inward along the cone, reaching $6.11M$ about $75M$ after $12M$ and $2.3M$ about $32M$ later (orange arrows). The black disk marks the horizon and the grey ring the excised region $r<\rin=2.3M$. For DRBH6, the gravitational acceleration differs from Schwarzschild by more than $10^{-3}$ only in the pink shell $2.3M<r<2.9M$, and the correction to the gravitational acceleration is lost to double-precision rounding beyond the dashed circle at $r = 4.65M$. Dotted circles mark the extraction radii $6.11M$ and $12M$, and the outer grey circle the accretion radius $r_{\rm acc}\simeq40M$.}
\label{fig:schematic}
\end{figure*}

Figure~\ref{fig:schematic} summarizes the structure of the flow and the regions relevant to our analysis. Gas arriving from the upstream side is focused into a shock cone whose stagnation point lies at $r_{\rm st}\simeq26.8M$. Inside the stagnation point, the post-shock gas falls toward the black hole and carries the variability inward. The Dymnikova correction changes the gravitational pull by more than $10^{-3}$ only in the thin shell $2.3M<r<2.9M$ (for DRBH6), and this correction to the acceleration falls below double-precision round-off beyond $r \simeq 4.65M$ (the lapse correction itself is already lost beyond $r_\epsilon = 4.49M$, Table~\ref{tab:models}). We first examine the flow structure and its local response to the core (Sec.~\ref{subsec:morphology}), then the time evolution and the sensitivity of the flow to perturbations (Sec.~\ref{subsec:chaos}), the time-averaged accretion rate (Sec.~\ref{subsec:mean}), the propagation of fluctuations (Sec.~\ref{subsec:lags}), and finally the power spectra (Sec.~\ref{subsec:psd}).

\subsection{Shock-cone morphology and the local response to the core}
\label{subsec:morphology}

\begin{figure*}[t]
\centering
\includegraphics[width=\textwidth]{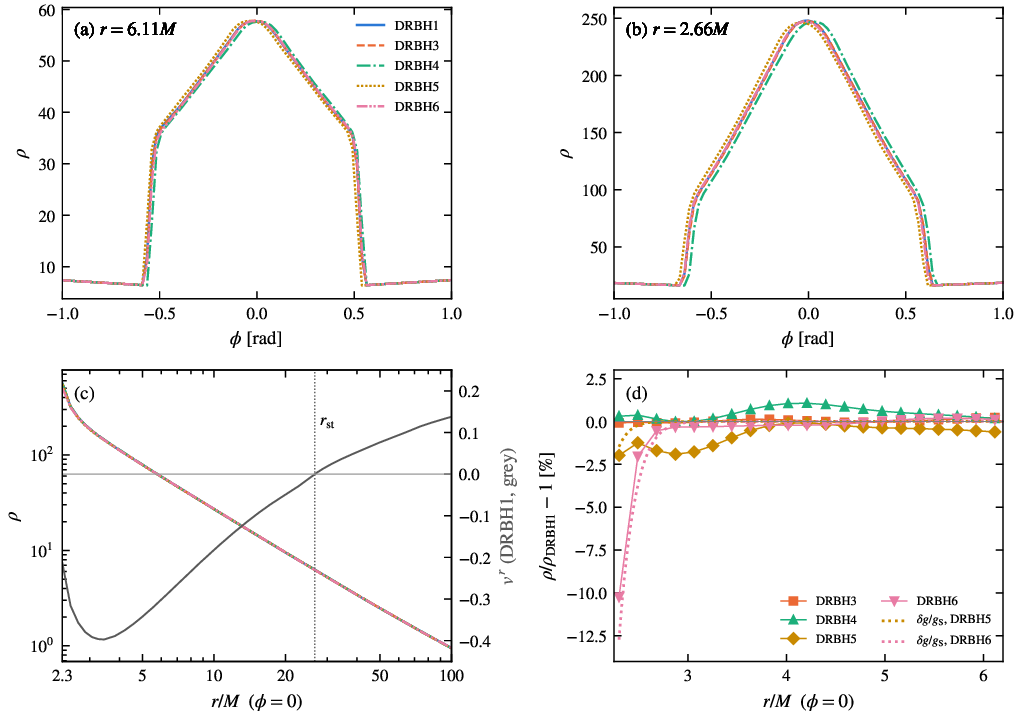}
\caption{Flow structure at the end of each run. (a), (b) Azimuthal density profiles at $r=6.11M$ and $2.66M$; the shock fronts bound the cone, whose axis is at $\phi=0$. (c) Density along the downstream axis $\phi=0$ (coloured lines, indistinguishable on this scale) and radial velocity of DRBH1 (grey, right axis); the vertical line marks the stagnation point $r_{\rm st}$. (d) Relative density difference with respect to DRBH1 along the downstream axis (symbols) compared with the deviation of the gravitational acceleration, Eq.~(\ref{eq:dg}), for DRBH5 and DRBH6 (dotted lines).}
\label{fig:profiles}
\end{figure*}

Figure~\ref{fig:profiles} shows the structure of the flow at the end of each run. The azimuthal density profiles at $r=6.11M$ and $2.66M$ [panels (a) and (b)] display the characteristic BHL shock cone~\cite{Donmez:2010sx,LoraClavijo2013}. Upstream of the shock, the density decreases smoothly from $\rho\simeq10.05$ at $\phi=\pi$ to $\rho\simeq6.4$ just outside the cone at $r=6.11M$. It then jumps across the shock fronts, located at $|\phi|\simeq0.54$~rad, by a factor of $5.5$--$5.8$, and reaches a maximum $\rho_{\max}\simeq57.7$ on the cone axis. At $r=2.66M$, closer to the black hole, the cone is wider and denser, with $\rho_{\max}\simeq247$.

The cone geometry is the same in all models (Table~\ref{tab:global}). The full opening angle at $r=6.11M$, measured between the points of maximum density gradient located with sub-zone (parabolic) interpolation, is $1.077\pm0.003$~rad ($61.7^\circ$), and at $2.66M$ it is $1.214\pm0.002$~rad; grid-point estimates scatter by about one azimuthal zone ($\pm0.02$~rad), so the cone width is the same in all models to within the resolution. The peak density agrees within $0.3\%$ at $6.11M$ and $0.7\%$ at $2.66M$. The only visible difference is a rigid azimuthal displacement of the cone, by at most $0.03$~rad, roughly one zone, which varies non-monotonically with $\lcr$. Since the problem is symmetric about $\phi=0$, such displacements are snapshots of the flip-flop motion of the cone at different phases~\cite{Foglizzo2005,Donmez2012}, not an effect of the geometry. Along the downstream axis [Fig.~\ref{fig:profiles}(c)], the gas falls toward the black hole inside the stagnation point and moves outward beyond it. The stagnation point, where $v^r$ changes sign, is at $r_{\rm st}=26.8\pm0.3M$ in all models, i.e., at about two thirds of $r_{\rm acc}$, in a region where all metrics are bitwise identical.

The de Sitter core does, however, leave a direct imprint close to the inner boundary. Figure~\ref{fig:profiles}(d) shows the relative density difference along the downstream axis with respect to DRBH1. For DRBH3 and DRBH4, whose gravitational acceleration differs by less than $2\times10^{-4}$ (Table~\ref{tab:models}), the differences stay within about $\pm1\%$, which measures the snapshot-to-snapshot variability of the flow. For DRBH6, the density in the innermost radial zone is lower by $10.2\%$ ($\rho=503.5$ instead of $561.0$), while the magnitude of the orthonormal radial velocity measured by the Eulerian observer, $v^{\hat r}=\sqrt{\gamma_{rr}}\,v^r$, is higher, $|v^{\hat r}| = 0.633$ instead of 0.592. This response follows the weakening of the gravitational pull, $\delta g/g_{\rm S}=-12.4\%$ at $\rin$, and decays with it ($-3.8\%$ at $2.49M$): at $r=2.49M$ the density difference is $-2.1\%$ and at $r\gtrsim2.7M$ it is lost in the variability of the flow. Because density and velocity change in opposite directions, the product $r^2\rho v^r$ along the axis decreases by only 2.0\%. For DRBH5, $\delta g/g_{\rm S}=-1.5\%$ at $\rin$, and its density difference, about $-2\%$ over $2.3M\le r\le3.1M$, cannot be distinguished from the variability. The post-shock gas near the axis is slow and nearly in pressure equilibrium, so a weaker pull compresses it less; this is why the density, rather than the mass flux, carries the signature of the core. The response is confined to the innermost one or two sampled zones, $r\lesssim2.5M$, and does not propagate outward. Because the largest deficit occurs in the zone adjacent to the outflow boundary, part of it may be shaped by the boundary treatment; resolving it would require a higher radial resolution and an inner boundary inside the horizon. The same local signature appears in the upstream flow: at $r=2.66M$ and $\phi=\pi$, where $\delta g/g_{\rm S}=-1.1\%$ for DRBH6, its density is lower by $0.33\%$, while in all other models it agrees with DRBH1 to within $1.2\times10^{-4}$.

\begin{table*}[t]
\caption{Global properties of the flow. Columns 2--4: time-averaged accretion rate normalized to DRBH1 at the three radii, with $1\sigma$ uncertainties ($t\in[5000,19300]M$). Columns 5--7: fractional rms variability. Columns 8--9: time lags between radii from the cross spectrum ($8$--$120$~Hz). Columns 10--11: full opening angle of the shock cone at $r=6.11M$ and stagnation radius on the downstream axis at the end of the run. DRBH2 is bitwise identical to DRBH1.}
\label{tab:global}
\setlength{\tabcolsep}{3.4pt}
\begin{ruledtabular}
\begin{tabular}{lcccccccccc}
 & \multicolumn{3}{c}{$\langle\Mdot\rangle/\langle\Mdot\rangle_{\rm DRBH1}$} & \multicolumn{3}{c}{$\sigma/\langle\Mdot\rangle$ [\%]} & \multicolumn{2}{c}{lag [$M$]} & & \\
Model & $2.3M$ & $6.11M$ & $12M$ & $2.3M$ & $6.11M$ & $12M$ & $12\to6.11M$ & $6.11\to2.3M$ & $\Delta\phi_{\rm sh}$ [rad] & $r_{\rm st}/M$ \\
\hline
DRBH1 & $1.0000\pm0.0014$ & $1.0000\pm0.0013$ & $1.0000\pm0.0014$ & $1.47$ & $1.01$ & $0.72$ & $75.5\pm0.7$ & $31.8\pm0.2$ & $1.080$ & $26.6$\\
DRBH3 & $0.9967\pm0.0016$ & $0.9980\pm0.0015$ & $0.9998\pm0.0011$ & $1.60$ & $1.11$ & $0.67$ & $74.4\pm0.6$ & $31.5\pm0.2$ & $1.077$ & $27.2$\\
DRBH4 & $1.0014\pm0.0023$ & $1.0021\pm0.0017$ & $1.0020\pm0.0012$ & $1.84$ & $1.21$ & $0.67$ & $76.7\pm0.7$ & $31.3\pm0.2$ & $1.074$ & $26.5$\\
DRBH5 & $0.9967\pm0.0022$ & $0.9971\pm0.0023$ & $0.9979\pm0.0020$ & $1.74$ & $1.23$ & $0.85$ & $74.5\pm0.7$ & $32.1\pm0.2$ & $1.075$ & $26.8$\\
DRBH6 & $0.9970\pm0.0019$ & $0.9988\pm0.0017$ & $0.9996\pm0.0013$ & $1.63$ & $1.11$ & $0.68$ & $76.2\pm0.7$ & $31.3\pm0.3$ & $1.077$ & $26.9$\\
\end{tabular}
\end{ruledtabular}
\end{table*}

\subsection{Evolution of the accretion rate and sensitivity to perturbations}
\label{subsec:chaos}

\begin{figure*}[t]
\centering
\includegraphics[width=\textwidth]{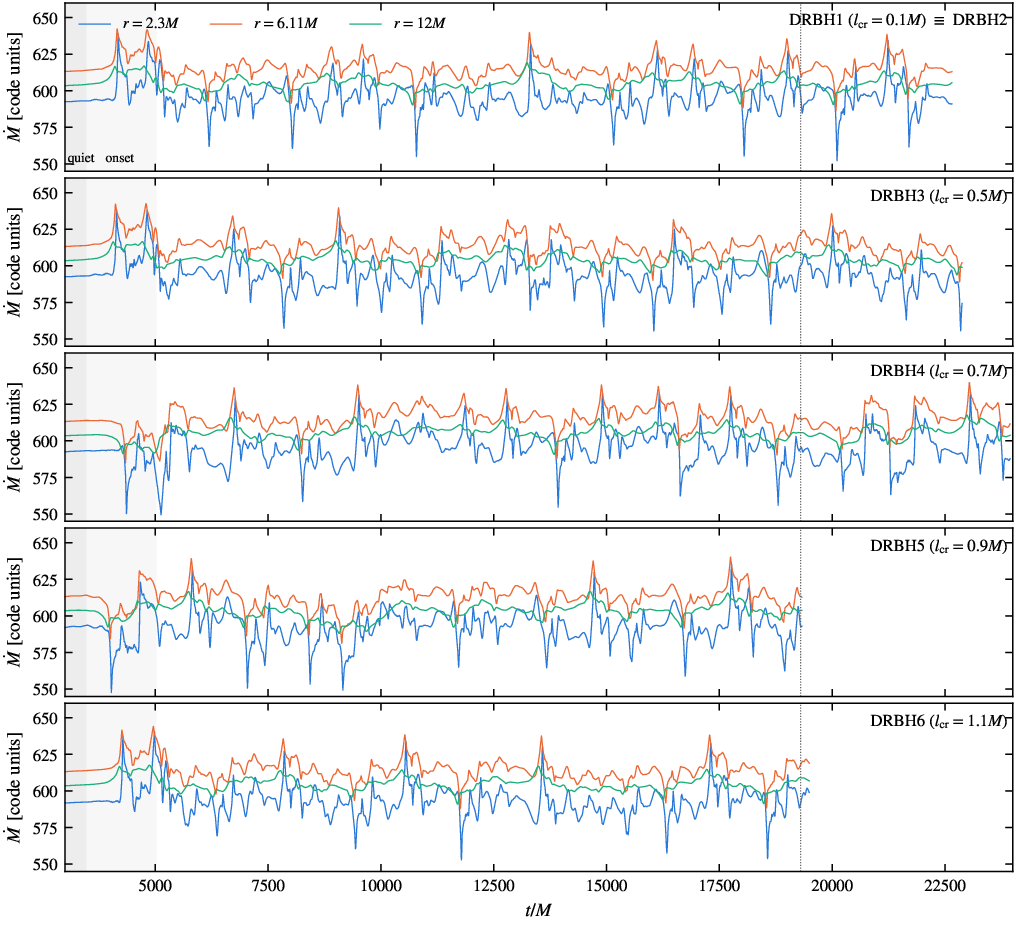}
\caption{Mass accretion rate at $r=2.3M$, $6.11M$, and $12M$ for the five independent runs, from $t=3000M$ to the end of each run ($5M$ bins). The darker and lighter grey bands mark the quiet phase and the onset of the instability; the vertical dotted line marks the end of the common analysis window, $t=19300M$. DRBH2 is bitwise identical to DRBH1.}
\label{fig:ts}
\end{figure*}

Figure~\ref{fig:ts} shows the accretion rate at the three radii for all independent runs. Until $t\simeq3500M$ the shock cone is nearly stationary and symmetric, and the accretion rate is constant to $\sigma/\langle\Mdot\rangle\simeq3\times10^{-4}$. At $t\simeq3550$--$3750M$ (measured at $\rin$) the cone becomes unstable, the oscillation amplitude grows with an $e$-folding time of $\simeq140$--$230M$, and it saturates at $t\simeq3950$--$4250M$. From $t\simeq5000M$ on, the flow is statistically stationary in all runs: the mean and the rms in consecutive windows of $3000$--$4000M$ agree within their uncertainties. The variability consists of sharp spikes and dips lasting $\sim50$--$100M$, superposed on slower modulations, and its amplitude grows toward the black hole: $\sigma/\langle\Mdot\rangle\simeq0.7\%$ at $12M$, $1.0$--$1.2\%$ at $6.11M$, and $1.5$--$1.8\%$ at $2.3M$, with excursions of up to $\pm7\%$ at $\rin$.

\begin{figure*}[t]
\centering
\includegraphics[width=\textwidth]{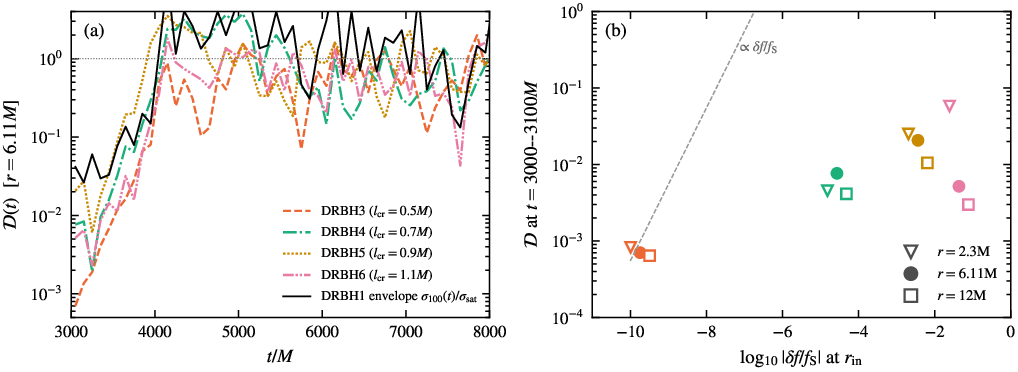}
\caption{(a) Normalized separation $\mathcal{D}(t)$, Eq.~(\ref{eq:sep}), between each run and DRBH1 at $r=6.11M$, together with the rms amplitude of DRBH1 in $100M$ windows normalized to its saturated value (black). (b) Separation during the quiet phase ($t=3000$--$3100M$) at the three radii as a function of the metric deviation at the inner boundary; the dashed line shows a separation proportional to $\delta f/f_{\rm S}$, normalized to DRBH3.}
\label{fig:decor}
\end{figure*}

The time series of DRBH1 and DRBH2 are bitwise identical over the whole evolution, which confirms that the correction of DRBH2 is lost to rounding inside the domain (Sec.~\ref{subsec:screening}). DRBH3, whose metric differs at $\rin$ by only $1.8\times10^{-10}$, follows DRBH1 closely during the quiet phase, with $\mathcal{D}\simeq10^{-3}$ at $t=3000M$, but the two runs separate exponentially during the onset and become fully decorrelated by $t\simeq4050M$ [Fig.~\ref{fig:decor}(a)]. The separation grows at a rate comparable to that of the amplitude of the instability itself (black line), with an $e$-folding time of $\simeq 150$--$300M$ (cf.\ $\simeq 140$--$230M$ for the amplitude), after which $\mathcal{D}$ fluctuates around unity, the value expected for independent signals. In the saturated phase, the correlation coefficient between DRBH1 and any other run, computed in $400M$ windows, averages $|\bar r|<0.12$. The same happens for DRBH4--DRBH6.

Defining the decorrelation time by $\mathcal{D}=0.5$, all runs decorrelate within $t_{\rm dec}\simeq3850$--$4850M$. Because $\mathcal{D}$ is normalized by the saturated rms, $t_{\rm dec}$ is necessarily tied to the saturation of the instability, so it is the seed separation before the onset that measures the sensitivity to the perturbation. Figure~\ref{fig:decor}(b) shows it against $\delta f/f_{\rm S}$ at $\rin$. While the metric perturbation grows by eight orders of magnitude from DRBH3 ($1.8\times10^{-10}$) to DRBH6 ($4.4\times10^{-2}$), the separation at $6.11M$ and $12M$ grows only from $\simeq7\times10^{-4}$ to $\sim10^{-2}$, and it is not even monotonic in $\delta f/f_{\rm S}$ (DRBH5 is more separated than DRBH6). The seed of the instability is therefore not the metric difference itself but the truncation-level noise of the evolution, which the smallest perturbations already raise to $\mathcal{D}\sim10^{-3}$ before $t=3000M$, as expected for a system with sensitive dependence on initial conditions~\cite{Lorenz1963,Ott2002}. At $\rin$, the separation of DRBH6 is larger ($\mathcal{D}\simeq0.06$) because it includes the systematic offset discussed below, and it grows more slowly than at the other radii. Runs whose metric differs at $\rin$ also have slightly different Courant-limited time steps, which adds a further truncation-level perturbation. Whatever its origin, a difference in the time series between two models carries no information about the geometry. The physically meaningful comparison is between statistical properties, and the scatter between DRBH1, DRBH3, and DRBH4, whose gravitational acceleration differs by less than $2\times10^{-4}$, gives the intrinsic uncertainty of each observable.

The quiet phase, before the instability develops, provides a clean test of the response of the accretion rate itself. For $3000M\le t\le3500M$, the accretion rate at $\rin$ of DRBH6 is lower than that of DRBH1 by $0.69$ ($1.2\times10^{-3}$ in relative terms, with an rms scatter of the difference of only $0.09$), while the difference for DRBH3 is $+0.02$ (scatter $0.05$). At $6.11M$ and $12M$, the relative differences for DRBH6 are below $10^{-4}$. This is the only systematic effect of the core on $\Mdot$ that we detect, and, consistent with Sec.~\ref{subsec:morphology}, it is confined to the inner boundary.

\subsection{Time-averaged accretion rate and variability amplitude}
\label{subsec:mean}

\begin{figure*}[t]
\centering
\includegraphics[width=\textwidth]{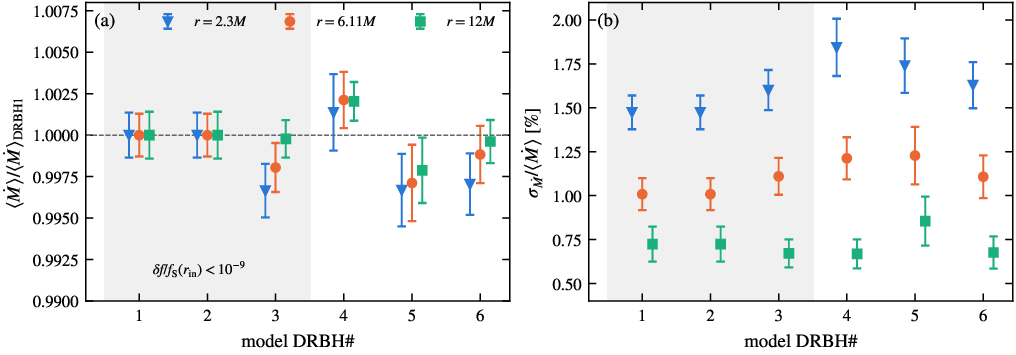}
\caption{(a) Time-averaged accretion rate normalized to DRBH1 and (b) fractional rms variability at the three radii, computed over $t\in[5000,19300]M$, with $1\sigma$ uncertainties that account for the autocorrelation of the series. The grey band marks the models whose metric equals Schwarzschild to better than $10^{-9}$ at $\rin$.}
\label{fig:means}
\end{figure*}

Figure~\ref{fig:means}(a) and Table~\ref{tab:global} give the time-averaged accretion rate of each model, normalized to DRBH1, at the three radii. All values lie within $0.33\%$ of unity, and none differs from DRBH1 by more than $1.7\sigma$ when the uncertainties of both runs are combined. There is no trend with $\lcr$: the largest negative deviations, $-0.33\%$ at $\rin$, are shared by DRBH3, whose geometry is Schwarzschild to $10^{-9}$, and DRBH5, while DRBH6 gives $-0.30\%$. Combining the measured offsets with their uncertainties, we bound the effect of the core on the mean accretion rate by $|\Delta\langle\Mdot\rangle|/\langle\Mdot\rangle<0.8\%$ (2$\sigma$) at all radii, even at $\rin$, where the gravitational pull of DRBH6 is weaker by $12\%$. The result reflects the fact that the mass supply is set by the BHL flow on the scale of $r_{\rm acc}\simeq40M$, where the metrics coincide, and that this supply is transmitted to the black hole regardless of the local geometry. Stationary spherical (Bondi) and thin-disk (Novikov--Thorne) models of accretion onto regular black holes, including the Dymnikova metric, have been studied in Ref.~\cite{Gambino2026}; the present simulations extend such studies to time-dependent, non-axisymmetric BHL flows.

The fractional rms [Fig.~\ref{fig:means}(b)] decreases outward in every model. At a given radius it varies between models by up to $25\%$ (from $1.47\%$ to $1.84\%$ at $\rin$), but not monotonically in $\lcr$: the largest values belong to DRBH4 and DRBH5, while DRBH6 is close to DRBH3. The skewness ($-0.38$ to $+0.30$ at $6.11M$) and excess kurtosis ($0.6$ to $2.1$) scatter in the same way, without order in $\lcr$, and the kurtosis range is already spanned by the Schwarzschild-equivalent runs DRBH1 and DRBH3.

\subsection{Inward propagation of accretion-rate fluctuations}
\label{subsec:lags}

\begin{figure*}[t]
\centering
\includegraphics[width=\textwidth]{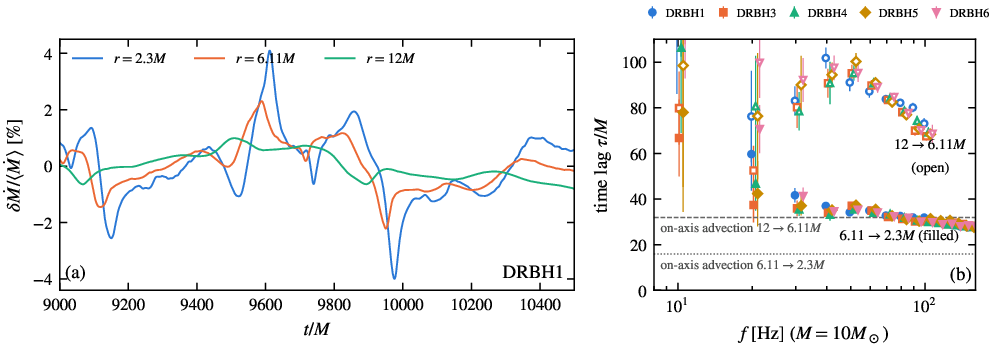}
\caption{(a) Fractional fluctuation of the accretion rate of DRBH1 at the three radii during $9000M\le t\le10500M$. (b) Time lag as a function of frequency between $12M$ and $6.11M$ (open symbols) and between $6.11M$ and $2.3M$ (filled symbols) for all independent runs. Horizontal lines mark the advection times along the downstream axis, $\int dr/(\alpha|v^r|)$, between the same radii.}
\label{fig:prop}
\end{figure*}

The variability at the three radii is strongly correlated, but it is not simultaneous. Figure~\ref{fig:prop}(a) shows a portion of the DRBH1 time series: each feature appears first at $12M$, then at $6.11M$, and finally at $2.3M$, where it is amplified. The cross spectra quantify this behaviour [Fig.~\ref{fig:prop}(b)]. Averaged over $8$--$120$~Hz, the coherence is $0.87$--$0.93$ between $12M$ and $6.11M$ and $0.95$--$0.98$ between $6.11M$ and $2.3M$; it is lower, $\simeq0.6$, at $10$--$20$~Hz, where the lags are poorly constrained. The time lags are positive at all frequencies and decrease with frequency: from $\simeq90$--$100M$ at $30$--$40$~Hz to $\simeq65M$ at $120$~Hz between $12M$ and $6.11M$, and from $\simeq35$--$40M$ to $\simeq29M$ between $6.11M$ and $2.3M$. The inverse-variance weighted mean lags, which are dominated by the well-measured higher frequencies, are $\tau_{12\to6.11}=75.5\pm1.0M$ and $\tau_{6.11\to2.3}=31.6\pm0.4M$, where the quoted uncertainty is the scatter between runs, larger than the statistical error of each run. The lags add up: the peak of the cross-correlation function between $12M$ and $2.3M$ is at $108\pm3M$.

A positive lag at all frequencies means that the fluctuations propagate inward, i.e., they are carried by the accreting gas rather than by sound waves moving outward. Their effective propagation speeds, $\simeq0.08$ between $12M$ and $6.11M$ and $\simeq0.12$ between $6.11M$ and $2.3M$, are lower than those of the gas on the cone axis: the coordinate advection times along $\phi=0$ are $32M$ and $16M$ (horizontal lines in Fig.~\ref{fig:prop}). The ratio of the measured lag to the on-axis advection time is $2.4$ for the outer interval and $2.0$ for the inner one. This indicates that the fluctuations are carried mainly by the slower gas that enters the cone through the shock fronts, where $|v^r|$ is about $40\%$ smaller than on the axis at $r=6.11M$ (azimuthal velocity profile at $r=6.11M$, not shown), and that they travel along longer, oblique paths. The steepening of the PSD between $12M$ and $2.3M$ discussed below shows that new high-frequency variability is added as the fluctuations approach the black hole.

The lags are the same in all models to within $1.6\%$ in both intervals (Table~\ref{tab:global}). DRBH6 gives $\tau_{6.11\to2.3}=31.3\pm0.3M$, compared with $31.65\pm0.2M$ for DRBH1 and DRBH3. The core modifies the geometry only in $r\lesssim2.5M$, which the gas crosses in the last few $M$ of its infall, and its effect on the propagation is therefore at most $\sim 1\%$, within the run-to-run scatter. More generally, the inward propagation shows that the variability measured near the black hole is generated at larger radii, in the cone and around the stagnation point, where the metric is exactly Schwarzschild, and is transported to the black hole by the flow. The local geometry near $\rin$ receives this signal but cannot shape it. This is the physical reason why the timing properties are insensitive to $\lcr$.

\subsection{Power spectra and the significance of QPOs}
\label{subsec:psd}

\begin{figure*}[t]
\centering
\includegraphics[width=\textwidth]{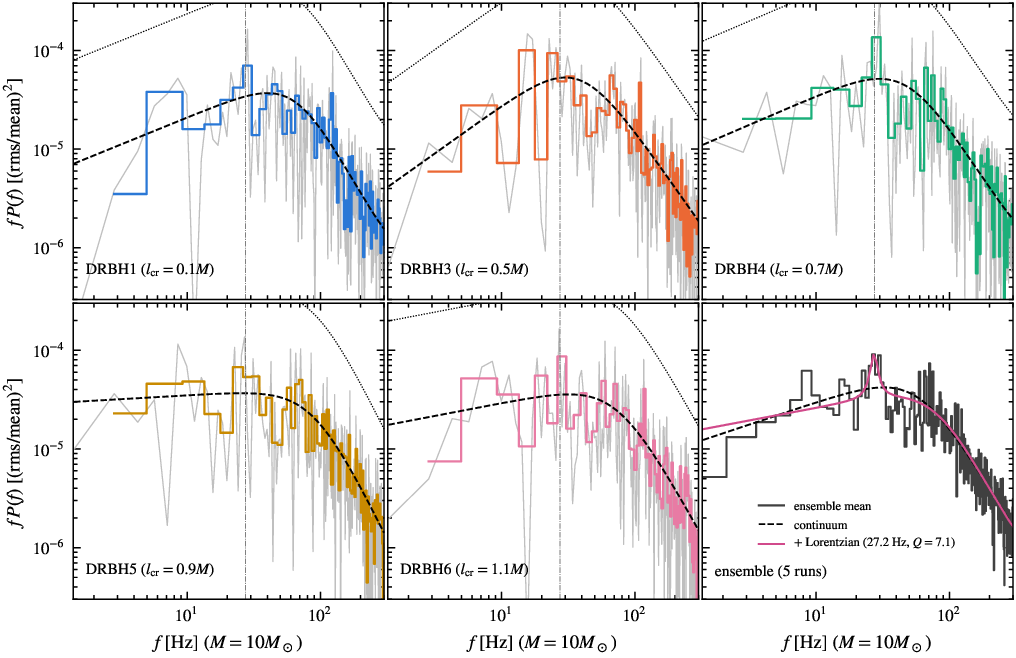}
\caption{Power spectra of the accretion rate at $r=6.11M$, in units of $(\text{rms}/\text{mean})^2$, for $M=10\Msun$. Grey: unbinned periodograms; coloured: averaged over three bins; dashed: best-fitting bending power law, Eq.~(\ref{eq:bpl}); dotted: global $99.73\%$ ($3\sigma$) threshold for a single unbinned bin. The vertical dash-dotted line marks $27.4$~Hz. The last panel shows the mean of the five independent periodograms with the continuum fit (dashed) and the continuum plus Lorentzian ($Q\ge2$) fit (solid).}
\label{fig:psd}
\end{figure*}

Figure~\ref{fig:psd} shows the PSDs at $r=6.11M$ in the $f\,P(f)$ representation, together with the best-fitting continuum, Eq.~(\ref{eq:bpl}). All spectra are broadband. The power rises slowly up to a break at $f_b \simeq 58$--$78$\,Hz in individual runs (for $M=10\Msun$) and falls steeply above it. The ensemble spectrum is described by $\alpha_1=0.55$, $\alpha_2=3.25$, and $f_b=58$~Hz at $6.11M$, corresponding to a time scale $1/f_b\simeq350M$. The spectral shape changes with radius in a systematic way: at $12M$ the low-frequency slope is steeper ($\alpha_1=1.64$) and the high-frequency cut-off sharper ($\alpha_2=6.0$), whereas at $2.3M$ the spectrum is flatter ($\alpha_1=0.32$, $\alpha_2=3.3$, $f_b=78$~Hz). The fractional rms in the $5$--$15$~Hz band is similar at the three radii ($0.4$--$0.8\%$), while in the $100$--$300$~Hz band it rises from $0.10\%$ at $12M$ to $0.27\%$ at $6.11M$ and $0.57\%$ at $2.3M$. The slow fluctuations are thus transported inward almost unchanged, while faster ones are generated or amplified in the inner flow.

No periodogram bin of any run reaches the global $3\sigma$ threshold at $6.11M$ (dotted lines in Fig.~\ref{fig:psd}), and the same holds at $2.3M$. The most significant narrow features (a single-bin spike at $123.5$~Hz in DRBH1 and a peak at $\simeq16$~Hz in DRBH3 at $12M$, with false-alarm probabilities $\simeq0.002$--$0.03$) appear in no other run and occur only in models whose geometry is Schwarzschild to $10^{-9}$ or better. Individual realizations of the same flow can therefore display features of $\simeq2$--$3\sigma$ significance by chance.

Lorentzian components with $Q\gg10$ fitted to single bins of individual runs are therefore not reliable QPOs: their widths are not resolved by the $1.42$~Hz resolution, and their significance against the red-noise continuum is low. The ensemble spectrum averages out such realization-specific features and reveals what is common to the flow. At $6.11M$ and $2.3M$, the QPO search ($Q\ge2$) on the ensemble spectrum finds a component at $f_0=27.2$ and $27.7$~Hz, with $Q=7.1$ and $7.3$ and fractional rms $0.36\%$ and $0.51\%$. The improvement of the fit is $\Delta C=11.0$ and $12.2$, which the simulations convert into false-alarm probabilities of $0.090$ and $0.048$ ($1.7\sigma$ and $2.0\sigma$); the $3\sigma$ threshold is $\Delta C\simeq16.5$. Without the $Q\ge2$ restriction, the best additional component at $6.11M$ is instead a broad bump ($f_0\simeq64$~Hz, $Q\simeq1.2$, $\Delta C=13.9$) that merely reshapes the continuum, whereas at $2.3M$ the $27.7$~Hz solution is recovered. At $12M$ the search converges to the boundary $Q=2$ ($f_0=9.9$~Hz, $\Delta C=24.5$, $p<0.002$), and with $Q$ free it gives $Q=1.9$. This is a peaked band-limited noise component that the single bending power law does not capture, rather than a QPO. A $27$--$28$~Hz feature, corresponding to a period of $\simeq740M$, thus recurs in the inner flow of independent realizations, but it is not statistically significant with the present data. Since $\Delta C$ of a real feature grows in proportion to the amount of data, a factor of $\simeq1.4$--$1.5$ more data, for example two to three additional realizations of the same length, would be needed to establish it at $3\sigma$.

Finally, we test directly whether the spectra depend on $\lcr$. Table~\ref{tab:psd} lists the median frequency of the variance and the band rms at $6.11M$. The band powers of the strongly corrected models DRBH5 and DRBH6 are consistent with those of the other runs: at $6.11M$ the smallest $p$ value of the $F$ test is $0.12$ for DRBH5 and $0.38$ for DRBH6, and at the other radii it is $0.026$ and $0.24$. Across all radii, bands, and models, the only $p<0.01$ ($p=0.003$) belongs to the $16$~Hz feature of DRBH3 noted above. The median frequency at $6.11M$ ranges from $12$ to $25$~Hz with uncertainties of $4$--$6$~Hz. It is lowest for DRBH5, but the difference from DRBH1 is only $1.5\sigma$, and DRBH6 ($18\pm6$\,Hz) is consistent with the Schwarzschild-equivalent runs within its uncertainty. At $2.3M$ and $12M$, $f_{50}$ ranges over $27$--$43$~Hz and $8$--$15$~Hz, respectively, again with DRBH5 and DRBH6 inside the range spanned by DRBH1, DRBH3, and DRBH4. The power spectra therefore carry no detectable signature of the de Sitter core.

\begin{table*}[t]
\caption{Spectral properties at $r=6.11M$ for $M=10\Msun$: median frequency $f_{50}$ of the variance between $1.4$ and $300$~Hz, fractional rms in four of the five bands used in the $F$ test (the 1.4--5\,Hz band is not shown), and the smallest two-sided $p$ value of that test against the other runs.}
\label{tab:psd}
\setlength{\tabcolsep}{8pt}
\begin{ruledtabular}
\begin{tabular}{lcccccc}
 & $f_{50}$ & \multicolumn{4}{c}{rms [\%] in band [Hz]} & \\
Model & [Hz] & 5--15 & 15--40 & 40--100 & 100--300 & $p_{\min}$ \\
\hline
DRBH1 & $25\pm6$ & $0.56\pm0.11$ & $0.58\pm0.07$ & $0.48\pm0.04$ & $0.27\pm0.02$ & 0.10\\
DRBH3 & $22\pm4$ & $0.47\pm0.09$ & $0.79\pm0.10$ & $0.46\pm0.04$ & $0.28\pm0.02$ & 0.07\\
DRBH4 & $20\pm5$ & $0.53\pm0.10$ & $0.73\pm0.09$ & $0.52\pm0.05$ & $0.25\pm0.02$ & 0.21\\
DRBH5 & $12\pm6$ & $0.70\pm0.13$ & $0.60\pm0.07$ & $0.48\pm0.04$ & $0.28\pm0.02$ & 0.12\\
DRBH6 & $18\pm6$ & $0.68\pm0.13$ & $0.61\pm0.07$ & $0.46\pm0.04$ & $0.27\pm0.02$ & 0.38\\
\hline
\multicolumn{7}{l}{Ensemble, all radii:}\\
$2.3M$ & \multicolumn{6}{l}{$f_b=78$~Hz; Lorentzian ($Q\ge2$) $27.7$~Hz, $Q=7.3$, rms $0.51\%$, $p=0.048$}\\
$6.11M$ & \multicolumn{6}{l}{$f_b=58$~Hz; Lorentzian ($Q\ge2$) $27.2$~Hz, $Q=7.1$, rms $0.36\%$, $p=0.090$}\\
$12M$ & \multicolumn{6}{l}{$f_b=127$~Hz; band-limited noise at $9.9$~Hz ($Q\simeq1.9$--$2$), not a QPO}\\
\end{tabular}
\end{ruledtabular}
\end{table*}

\section{Astrophysical implications}
\label{sec:astro}

\subsection{Timing properties of the shock cone and observed variability}
\label{subsec:xrb}

Since the timing properties do not depend on $\lcr$, comparing them with observations tests the shock-cone mechanism itself, which operates identically around Schwarzschild and Dymnikova black holes. For fixed dimensionless flow parameters, all time scales scale as $GM/c^3$, so the results can be transferred to any black-hole mass. Table~\ref{tab:scaling} lists the main time scales for several sources.

\begin{table*}[t]
\caption{Characteristic time scales of the simulated variability rescaled to astrophysical black holes: inward propagation lag between $6.11M$ and $2.3M$ ($31.6M$), break time scale of the ensemble PSD at $2.3M$ ($1/f_b=260M$), and period of the tentative $27$~Hz feature ($740M$). Masses from Refs.~\cite{EHT2019M87VI,Reid2014,Remillard2006,GRAVITY2022,Gierlinski2008}.}
\label{tab:scaling}
\setlength{\tabcolsep}{6pt}
\begin{ruledtabular}
\begin{tabular}{lccccc}
Source & $M\,[\Msun]$ & $GM/c^3$ & lag $6.11\to2.3M$ & $1/f_b$ ($f_b$) & tentative feature \\
\hline
$10\Msun$ reference & 10 & $49\,\mu$s & 1.6~ms & 13~ms (78~Hz) & 36~ms (27~Hz)\\
GRS~1915$+$105 & 12.4 & $61\,\mu$s & 1.9~ms & 16~ms (63~Hz) & 45~ms (22~Hz)\\
GRO~J1655$-$40 & 6.3 & $31\,\mu$s & 1.0~ms & 8~ms (124~Hz) & 23~ms (43~Hz)\\
Sgr~A$^*$ & $4.3\times10^6$ & 21~s & 11~min & 1.5~h & 4.4~h\\
RE~J1034$+$396 & $\sim2\times10^6$ & 9.9~s & 5~min & 43~min & 2.0~h\\
M87$^*$ & $6.5\times10^9$ & 8.9~h & 12~d & 96~d & 270~d\\
\end{tabular}
\end{ruledtabular}
\end{table*}

\subsubsection{Broadband noise} 
The accretion-rate variability of the shock cone is predominantly broadband, with a break at $f_b\simeq60$--$80$~Hz for $M=10\Msun$. In black-hole X-ray binaries, the broadband noise of the hard and intermediate states is decomposed into a few Lorentzian components with characteristic frequencies from $\sim0.1$ to several tens of Hz~\cite{Belloni2002}; the simulated break lies at the upper end of this range. The fractional variability of $\Mdot$, $0.7$--$1.8\%$, cannot be compared directly with the observed X-ray rms, which can reach tens of percent, without a radiative model linking $\Mdot$ to the emitted flux, but it shows that a non-rotating black hole fed by a low-angular-momentum flow produces broadband noise without any disk.

\subsubsection{Propagating fluctuations and time lags} 
The inward propagation of fluctuations measured in Sec.~\ref{subsec:lags} is the hydrodynamic analogue of the propagating-fluctuation picture used to explain the broadband noise and the hard time lags of X-ray binaries~\cite{Lyubarskii1997,Kotov2001,ArevaloUttley2006,Uttley2014}. In that picture, fluctuations generated at different radii of a disk propagate inward on the viscous time scale, which produces lags that decrease with frequency. In the shock cone, the propagation is advective and much faster: the lag between $6.11M$ and $2.3M$ is $1.6$~ms for $M=10\Msun$, and it also decreases with frequency, by about $25\%$ between $30$ and $120$~Hz. If the emission from the inner flow is harder than that from larger radii, this would produce hard lags of a few ms at tens of Hz. To our knowledge, such propagation lags have not been measured before in relativistic BHL simulations; we find that they are unaffected by the near-horizon geometry.

\subsubsection{QPOs}
The tentative $\simeq27$~Hz feature falls within the range of type-C low-frequency QPOs, which are usually found below $\sim10$~Hz but occasionally reach $\sim30$~Hz~\cite{Casella2005,Motta2016,IngramMotta2019}. Shock-cone oscillations could thus generate low-frequency QPOs around non-rotating black holes, without Lense--Thirring precession~\cite{IngramDoneFragile2009}, as in shock-oscillation models~\cite{Molteni1996}. The microquasar GRS~1915$+$105 shows a family of QPOs at $27$, $34$, $41$, and $67$~Hz~\cite{Morgan1997,Strohmayer2001,BelloniAltamirano2013,MottaBelloni2024}. For its dynamical mass, $M=12.4^{+2.0}_{-1.8}\Msun$~\cite{Reid2014}, the simulated feature corresponds to $19$--$26$~Hz, close to the lowest of these frequencies. However, given the low significance of the simulated feature, the uncertainty of the mass, and the dependence of the cone frequencies on $V_\infty$ and on the position of the stagnation point~\cite{Donmez2022Universe,Donmez2024JCAP}, this agreement is a consistency check rather than an identification. The high-frequency QPO pairs with a $3{:}2$ ratio, observed at $2\nu_0$ and $3\nu_0$ with $\nu_0 \simeq 931\,(M/M_\odot)^{-1}$\,Hz~\cite{Remillard2006}, i.e., at $\simeq190$ and $280$~Hz for $10\Msun$, lie well above the break of the simulated spectra, where the shock-cone variability carries little power. They must originate closer to the black hole, for instance from epicyclic resonances~\cite{AbramowiczKluzniak2001,Torok2005} or oscillating tori~\cite{Rezzolla2003}. Since the epicyclic frequencies are set at $r\gtrsim6M$, where $|\delta g/g_{\rm S}|<1.5\times10^{-34}$ [Eq.~(\ref{eq:bounds})], those QPOs are equally blind to a Dymnikova core.

\subsubsection{Supermassive black holes} 
Sgr~A$^*$ is fed by the winds of massive stars in the central parsec, which produce a low-angular-momentum inflow~\cite{Ressler2018}, a situation in which BHL-type accretion is a reasonable approximation. For its mass, the simulated variability would appear as broadband fluctuations with a break at $\simeq1.5$~h, propagation lags of $\sim10$~min between the inner radii, and a possible modulation at $\simeq4$~h. The observed quasi-periodicities of the near-infrared flares, $\simeq17$~min~\cite{Genzel2003}, and the $\simeq70$~min period of the polarization loop after an X-ray flare~\cite{Wielgus2022} are shorter and are naturally explained by orbital motion within a few Schwarzschild radii. Shock-cone variability would therefore contribute slower, hour-scale modulations that could be searched for in long multiwavelength light curves. The $\simeq1$~h QPO of RE~J1034$+$396~\cite{Gierlinski2008} lies between the break and the tentative feature for $M\sim2\times10^6\Msun$, but its host accretes through a disk. For M87$^*$, the corresponding time scales, months to a year, are accessible to multi-year VLBI monitoring~\cite{EHT2019M87VI,Cui2023}. Two caveats apply. The asymptotic velocity $V_\infty=0.2c$ adopted here is much larger than typical velocities in galactic nuclei, so the rescaled time scales are order-of-magnitude estimates. Moreover, if $\lcr$ is a fixed microscopic length, as suggested by its RG origin~\cite{Platania:2019kyx}, then $\lcr/M$ is many orders of magnitude smaller for astrophysical black holes than in our models, and the core is even more deeply screened.

\subsection{Can accretion reveal a de Sitter core?}
\label{subsec:signatures}

Our results answer this question quantitatively, and Table~\ref{tab:summary} collects them. Within a BHL flow onto any Dymnikova black hole, the core can only affect the gas inside $r\simeq3M$. There, the weakening of the gravitational pull produces a local density deficit that scales with $\delta g/g_{\rm S}$ and does not propagate outward, because the variability is generated at large radii and carried inward. Global observables, namely the mean accretion rate, the variability amplitude, the cone geometry, the propagation lags, and the power spectrum, differ between models by less than the scatter of independent realizations. For the strongest correction we simulated, $\lcr=1.1M$, which is already $97\%$ of the extremal value, we obtain the bounds $|\Delta\langle\Mdot\rangle|/\langle\Mdot\rangle<0.8\%$ (2$\sigma$) and $|\Delta\tau|/\tau<2\%$, and band rms values consistent with those of the Schwarzschild-equivalent runs within their $7$--$20\%$ uncertainties.

\begin{table*}[t]
\caption{Which observables respond to the de Sitter core? For each quantity, the range spanned by the three Schwarzschild-equivalent runs (DRBH1, DRBH3, DRBH4; $|\delta g/g_{\rm S}|<2\times10^{-4}$ at $\rin$) is compared with the values for the two strongly corrected models. Only the two local quantities at the inner boundary (axial density and quiet-phase $\dot M$) depart from the Schwarzschild-equivalent range by more than the realization scatter.}
\label{tab:summary}
\setlength{\tabcolsep}{6pt}
\begin{ruledtabular}
\begin{tabular}{lcccc}
Observable & DRBH1, 3, 4 (range) & DRBH5 & DRBH6 & Responds to core? \\
\hline
$\delta g/g_{\rm S}$ at $\rin$ & $<2\times10^{-4}$ & $-1.5\%$ & $-12.4\%$ & --- \\
Axial density at $\rin$ & $560.6$--$562.7$ & $549.9$ & $503.5$ & yes (local) \\
Quiet-phase $\Mdot(\rin)$, offset from DRBH1 [$10^{-4}$] & $0$ to $+1.9$ & $-0.7$ & $-11.7$ & yes (local) \\
$\langle\Mdot\rangle/\langle\Mdot\rangle_{\rm DRBH1}$ at $2.3M$ & $0.9967$--$1.0014$ & $0.9967$ & $0.9970$ & no \\
$\langle\Mdot\rangle/\langle\Mdot\rangle_{\rm DRBH1}$ at $12M$ & $0.9998$--$1.0020$ & $0.9979$ & $0.9996$ & no \\
$\sigma/\langle\Mdot\rangle$ at $2.3M$ [\%] & $1.47$--$1.84$ & $1.74$ & $1.63$ & no \\
Lag $12\to6.11M$ [$M$] & $74.4$--$76.7$ & $74.5$ & $76.2$ & no \\
Lag $6.11\to2.3M$ [$M$] & $31.3$--$31.8$ & $32.1$ & $31.3$ & no (within $1.6\%$) \\
Cone opening angle at $6.11M$ [rad] & $1.074$--$1.080$ & $1.075$ & $1.077$ & no \\
Stagnation radius $r_{\rm st}$ [$M$] & $26.5$--$27.2$ & $26.8$ & $26.9$ & no \\
$f_{50}$ at $6.11M$ [Hz] & $20$--$25$ ($\pm5$) & $12\pm6$ & $18\pm6$ & no \\
rms in $100$--$300$~Hz at $6.11M$ [\%] & $0.25$--$0.28$ & $0.28$ & $0.27$ & no \\
\end{tabular}
\end{ruledtabular}
\end{table*}

This contrasts with other modified spacetimes studied with the same method. For the hairy Horndeski black hole, whose correction decays only logarithmically, the hair parameter moves the stagnation point and changes the accretion rate substantially~\cite{Donmez2024JCAP}. For the Kerr metric in the infrared limit of asymptotic safety, with corrections $\propto\xi/r^2$, the accretion rate changes noticeably near the horizon~\cite{Donmez2026JCAP}. The sensitivity of BHL accretion to a modified geometry is thus determined by how far the deviation extends beyond the horizon. The Dymnikova correction, screened as $e^{-r^3/(2\lcr^2M)}$, is the extreme case. The same screening explains why the fundamental quasinormal modes of the Dymnikova black hole are almost unaffected, while overtones, which probe the geometry closer to the horizon, respond more strongly~\cite{Konoplya:2023aph,Lutfuoglu2026AnnPhys}. For the same reason, its grey-body factors and Hawking emission differ from Schwarzschild mainly through the modified Hawking temperature~\cite{Dubinsky:2025nxv,Skvortsova:2026mwz}.

Accretion can therefore reveal a Dymnikova core only under conditions that differ from those of an astrophysical black hole with a microscopic $\lcr$. Spin may also enlarge the region in which the correction is felt, since rotation deforms the de Sitter core into a de Sitter disk~\cite{DymnikovaGalaktionov2015}; BHL accretion onto rotating Dymnikova black holes is left for future work. One possibility is the near-extremal and horizonless regime, $\lcr\gtrsim\lcr^{\rm ext}$, where the flow can reach the regular core. For horizonless configurations of other modified spacetimes, the BHL flow is markedly quieter~\cite{Donmez2026JCAP}. Resolving the region between $r_+$ and $3M$ would also require horizon-penetrating coordinates and an inner boundary inside the outer horizon $r_+$. Any proposed signature must, in addition, exceed the realization-to-realization scatter identified here. This calls for ensembles of simulations, which we have shown to be the proper basis for comparing spacetimes through nonlinear accretion flows.

%
\section{Conclusions}
\label{sec:conclusions}

We have performed general relativistic hydrodynamic simulations of BHL accretion onto the Dymnikova regular black hole for six values of the critical length, $0.1M\le\lcr\le1.1M$, and analyzed the mass accretion rate at three radii with methods that separate physical effects from the intrinsic variability of the flow. Our main conclusions are:

\begin{enumerate}
\item The Dymnikova correction is screened as $e^{-r^3/(2\lcr^2M)}$. For every Dymnikova black hole, the relative deviation of the lapse is below $6\times10^{-5}$ at the photon sphere and below $3\times10^{-37}$ at the ISCO [Eq.~(\ref{eq:bounds})]. The dynamically relevant quantity, the gravitational acceleration, is reduced more strongly, by up to $12\%$ at $r=2.3M$ for $\lcr=1.1M$, owing to the factor $1+3x$ in Eq.~(\ref{eq:dg}).

\item The flow responds to the weaker gravity only locally: for $\lcr=1.1M$, the post-shock density on the downstream axis decreases by $10\%$ in the innermost radial zone, the infall speed increases, and the accretion rate at $2.3M$ is lower by $0.12\%$ during the quiet phase. The response follows $\delta g/g_{\rm S}$ and does not reach $r\gtrsim2.7M$.

\item The shock cone becomes unstable at $t\simeq3600M$ and saturates at $t\simeq4000M$. Runs whose metrics differ by as little as $1.8\times10^{-10}$ decorrelate completely during the saturation, and the seed separation before the onset grows far more slowly than the metric perturbation, showing that the instability is seeded by truncation-level noise. Time series of different models can therefore only be compared statistically, and the scatter between nearly identical geometries sets the intrinsic uncertainty.

\item The time-averaged accretion rate is independent of $\lcr$ to within $0.8\%$ (2$\sigma$) at all radii, and the variability amplitude, skewness, kurtosis, shock-cone opening angle ($1.077\pm0.003$~rad at $6.11M$), and stagnation point ($26.8\pm0.3M$) vary between models only within the realization scatter.

\item Accretion-rate fluctuations propagate inward with weighted mean lags of $75.5\pm1.0M$ ($12\to6.11M$) and $31.6\pm0.4M$ ($6.11\to2.3M$), with frequency-averaged coherence above $0.87$ and the same values for all $\lcr$ to within $1.6\%$. The variability is generated at large radii, where the geometry is Schwarzschild, and transported to the black hole, which explains the insensitivity of the timing properties to the core.

\item The power spectra are broadband, with a break at $f_b\simeq60$--$80$~Hz for $M=10\Msun$ in the inner flow, and high-frequency power grows toward the black hole. No individual run shows a QPO above $3\sigma$ once the red-noise continuum is taken into account. A feature at $\simeq27$~Hz ($Q\simeq7$) recurs in the ensemble spectrum of the inner flow at $1.7$--$2.0\sigma$; about $1.4$--$1.5$ times more data would be needed to confirm it at $3\sigma$. The band powers of the strongly corrected models are consistent with those of the Schwarzschild-equivalent runs.
\end{enumerate}

In summary, BHL accretion onto a Dymnikova black hole is observationally indistinguishable from accretion onto a Schwarzschild black hole. Timing observations of X-ray binaries and supermassive black holes therefore test the shock-cone mechanism, not the regular core. Beyond the Dymnikova case, our analysis shows that statistical tests based on independent realizations are needed before differences between simulated accretion flows in different spacetimes are interpreted as physical signatures. Future work will address horizonless configurations, three-dimensional flows, and ensembles of realizations long enough to establish the shock-cone QPO spectrum at high significance.


\section*{Acknowledgments}

All numerical simulations were performed using the Phoenix High Performance Computing facility at the American University of the Middle East (AUM), Kuwait.  

\section*{Data Availability Statement}

The data sets generated and analyzed during the current study were produced using high-performance computing resources. These data are not publicly available due to their large size and computational nature, but are available from the corresponding author upon reasonable request.

\bibliographystyle{apsrev4-1}
\bibliography{references}
\end{document}